\documentclass[twocolumn,showpacs,preprintnumbers,amsmath,amssymb]{revtex4}

\usepackage{graphicx}
\usepackage{dcolumn}
\usepackage{bm}

\newcommand{\ket} [1] {\vert #1 \rangle}
\newcommand{\bra} [1] {\langle #1 \vert}

\begin{document}


\title{Experimental quantum communication complexity}

\author{Pavel Trojek$^{1,2}$}
\author{Christian Schmid$^{1,2}$}
\author{Mohamed Bourennane$^{1,2}$}
\author{\v{C}aslav Brukner$^3$}
\author{Marek \.{Z}ukowski$^4$}
\author{Harald Weinfurter$^{1,2}$}
\affiliation{$^1$ Max-Planck-Institut f\"{u}r Quantenoptik,
D-85748 Garching, Germany
\\
$^2$ Sektion Physik,
Ludwig-Maximilians-Universit\"{a}t, D-80799 M\"{u}nchen, Germany
\\
$^3$ Institut f\"{u}r Experimentalphysik, Universit\"{a}t Wien,
Boltzmanngasse 5, A-1090, Wien, Austria
\\
$^4$ Instytut Fizyki
Teoretycznej i Astrofizyki Uniwersytet Gda\'{n}ski, PL-80-952
Gda\'{n}sk, Poland}

\date{June 23, 2004}

\begin{abstract}
We prove that the fidelity of two exemplary communication
complexity protocols, allowing for an N-1 bit communication, can
be exponentially improved by N-1 (unentangled) qubit
communication. Taking into account, for a fair comparison, all
inefficiencies of state-of-the-art set-up, the experimental
implementation outperforms the best classical protocol, making it
the candidate for multi-party quantum communication applications.
\end{abstract}

\maketitle

Quantum information science transgresses limitations of
conventional information transfer, cryptography and computation.
Recently, significant advantages were recognized when applying
quantum phenomena in the field of communication complexity
problems (CCP's) \cite{yao79}. There, separated parties,
performing {\em local} computations, exchange information in order
to accomplish some {\em globally} defined task. Two types of CCP's
are distinguished: the first minimizes the amount of information
exchange necessary to solve the task with certainty
\cite{cleve97,buhrman99,buhrman97}. The second maximizes the
probability of successfully solving the task for restricted amount
of information \cite{buhrman97,hardy99,brukner02}. Such studies
aim, e.g., at a speed up of distributed computations by increasing
the communication efficiency, or at an optimization of VLSI
circuits and data structures \cite{kushilevitz97}.

Quantum CCP protocols, using multi-particle entanglement, were
proven to be clearly superior with respect to classical ones
\cite{buhrman97,buhrman99,cleve97,hardy99,brukner02}. However, the
technology of entanglement based multi-party quantum communication
is still in a premature stage. A recent reformulation of quantum
CCP's pointed out that even the communication employing single
qubits may outperform classical CCP's
\cite{buhrman98,raz99,galvao01}. Such a simplification would be of
tremendous importance, as it would make a multi-party
communication task technologically comparable to quantum key
distribution, the only commercial application of quantum
information science so far.

Here we prove that, for CCP's with restricted communication, the
superiority of the single qubit assisted protocols over the
corresponding classical ones may increase even exponentially with
the number of partners. Furthermore, using parametric
down-conversion as a source of heralded single qubits, we
experimentally show that quantum protocols solve two exemplary
CCP's more efficiently, even with the limited detection efficiency
inherent in real single-photon experiments. By solving these CCP's
with a sequential transfer of a single qubit only, we demonstrate
a generic way of bringing multi-party quantum communication
schemes much closer to realistic applications.

Let us first introduce the two CCP's analyzed and implemented
here. The first one, problem A, is the so called {\em modulo-4
sum} problem \cite{buhrman97,buhrman99,galvao01}. Imagine $N$
separated partners $P_1, \ldots, P_N$. Each of them receives a
two-bit string $X_k$, $\left(X_k=0,1,2,3; k = 1,\ldots,N \right)$.
The $X_k$'s are distributed such that their sum is even, i.e.
$(\sum_{k=1}^{N} X_k){\textrm{mod} 2} = 0$. No partner has any
information whatsoever on the values received by the others. The
partners then communicate with the common goal that one of them,
say $P_N$, can tell whether the sum modulo-4 of all input strings
is equal 0 or 2. That is, $P_N$ announces the value of the
dichotomic, i.e. equal $\pm1$, function $T(X_1,...,X_N)$ given by
$T_A(X_1,\ldots,X_N) = 1-(\sum_{k=1}^{N} X_k)\textrm{mod} 4$ (for
an alternative formulation see footnote \cite{remark1}). The
partners can freely choose how to communicate information about
their $X_k$, i.e. they can choose between sequential communication
 from one to the other or any arbitrary tree-like
structure ending at the last party $P_N$. However, the total
amount of communication is restricted to only $N-1$ bits
(classical scenario).

Problem B  has a similar structure, but now $N$ {\em real} numbers
$X_1,\ldots,X_N \in [0,2 \pi)$ with probability density
\begin{equation}\label{distrrho}
 p_B(X_1,\ldots,X_N) = \frac{1}{4(2\pi)^{N-1}}|\cos(X_1+ \ldots +X_N)|
\end{equation}
are distributed to the partners. Their task is to compute whether
$ \cos(X_1+ \ldots +X_N)$ is positive or negative, i.e. to give
the value of the function $T_B= S[ \cos (\sum_{k=1}^{N} X_k)]$,
where $S(x)=x/|x|$. The communication restriction is the same as
for problem A.

To find the best performing classical protocols for these CCP's,
we first rewrite the random inputs $X_k$. For the task A we put
$X_k = (1-y_k) + x_k$, where $y_k \in \{-1,1\}$, $x_k \in
\{0,1\}$. For the task B we write $X_k = \pi (1-y_k)/2 + x_k$,
with $y_k \in \{-1,1\}$, $x_k \in [0,\pi)$. Note that the
dichotomic variables $y_k$ are not restricted by the probability
distributions for the $X_k$'s. Thus they are completely random.
The global task function $T$ can now be put as $T=\prod_{k=1}^N
y_k f(x_1,...,x_N)$, see \cite{reformulation}, and
$p(X_1,...,X_N)=2^{-N}p'(x_1,...,x_N)$.

Depending on the value of the product of all $y_k$'s the value of
$T$ flips between $\pm1$. Thus, if information on $y_k$ of any of
the partners is omitted in the course of the protocol, the result
is completely random. This implies that all $N$ partners must be
involved in an unbroken communication structure. Because of  the
restriction to maximum $N-1$ bits of communication, each of the
partners must send only {\em one} bit, except for the last one,
$P_N$, who gives the result \cite{broadcasting}.

Each of these one-bit messages encodes the value of a dichotomic
function $e_k=\pm1$. It depends on the local input number $X_k$
and possibly on information $\{e_l,e_m,\ldots\}$ already received
from other partners. Due to the highly restricted form of two
valued functions, see \cite{dichotomic}, one can express any $e_k$
in the form $e_k = b_k(x_k, e_l, e_m, \ldots) + c_k(x_k, e_l, e_m,
\ldots)y_k$. In order to obtain a non-random final result, $e_k$
{\it must} depend on $y_k$. Thus, $b_k$ must be equal 0, while
$c_k$ itself is now a dichotomic function. Continuing the
expansion of $c_k(x_k, e_l, e_m, \ldots)$ and keeping in mind that
all previous messages received by the $k$-th partner must be taken
into account, we obtain $e_k = y_k a_k(x_k) e_l e_m \ldots $,
where $a_k$ is again a dichotomic function depending only on the
local input $x_k$. Next, one expands in a similar way $e_l, e_m,
\ldots$, which leads to
$e_k=y_ka_k(x_k)y_la_l(x_l)y_ma_m(x_m)\ldots$. The final answer,
$e_N$, given by $P_N$ must have the same structure as $e_k$ and
therefore, it must be equal to $e_N = \prod_{i=1}^{N}a_i(x_i)y_i$,
\cite{sequential}.

If the answer is correct, then $T = e_N$, and thus $T\cdot e_N =
1$. Otherwise, $T \cdot e_N = -1$. Thus, one can introduce the
measure $F$ (fidelity) of the average success in the form
\begin{equation}
F_c =\left| \sum\limits _{X_1, \ldots, X_N}p(X_1, \ldots,
X_N)T(X_1, \ldots, X_N)e_N \right|.
\end{equation}
For the problem B the summations are replaced by integrations. The
probability of success, $P$, is given by $P =(1+F)/2$. For the
best classical protocols of the CCP's given above we obtain
\begin{eqnarray}
F_c&=&\left|2^{-N}\sum\limits _{x_1, \ldots, x_N}\sum\limits
_{y_1, \ldots, y_N=\pm1} p'(x_1, \ldots, x_{N}) \prod_{l=1}^N
y_l\right.
\nonumber\\
 && \left. \times f(x_1, \ldots, x_{N}) e_N(X_1,...,X_N) \right|
\nonumber \\
&=&\left|\mathop{\sum}\limits _{x_1, \ldots, x_N}g(x_1, \ldots,
x_N)a_1(x_1)\ldots a_N(x_N)\right|,
\end{eqnarray}
where we denoted the product $p'(x_1, \ldots, x_{N}) f(x_1,
\ldots, x_{N})$ by $g(x_1, \ldots, x_{N})$. Since $F_c$ depends on
the product of local functions $|a_i(x_i)|\leq1$, it is bounded
from above, i.e., $F_c\leq B(N)$ \cite{brukner02}.

The  bounds $B(N)$ for our problems A and B can be easily
calculated. In both cases the fidelity decreases exponentially
with number $N$ of parties. For task A one has $F_{c,A}=2^{-K+1}$,
where $K=N/2$ and $K=(N+1)/2$ for even and odd number of parties,
respectively. This {\em analytic} result, valid for arbitrary $N$,
confirms the numerical simulations of \cite{galvao01} for small
$N$. For task B we derived $F_{c,B}=(2/\pi)^{N-1}$. Due to the
formal analogies the integrals needed to get this result already
appeared in the derivation of a Bell inequality involving
continuous range of settings \cite{zukowski93}.

The Holevo bound \cite{holevo73} limits the information storage
capacity of a qubit to exactly one classical bit. Thus, we
restrict the maximum communication exchange for quantum protocols
of the presented CCP's to $N-1$ qubits, or alternatively, to
$N-1$-times exchange of a {\em single} qubit. The solution of task
A  starts with a qubit in the state
$\ket{\psi_i}=2^{-1/2}(\ket{0}+\ket{1})$. Parties sequentially act
on the qubit with the unitary phase-shift transformation of the
form $ \ket{0}\bra{0} + e^{i\pi/2X_k}\ket{1}\bra{1}$, in
accordance with their local data. After all $N$ phase shifts the
state is
\begin{equation}
\ket{\psi_f} = \frac{1}{\sqrt{2}}(\ket{0}+e^{i\pi/2(\sum_{k=1}^{N}
X_k)}\ket{1}).
\end{equation}
Since the sum over $X_k$ is even, the phase factor
$e^{i\pi/2(\sum_{k=1}^{N} X_k)}$ is equal to the dichotomic
function $T_A$ to be computed. Therefore, a measurement of the
qubit in the basis $ (\ket{0}\pm \ket{1})/\sqrt{2}$ reveals the
value of $T_A$ with fidelity $F_{q,A} = 1$, that is, always
correctly.

Our quantum protocol for task B starts with a qubit in the same state $
\ket{\psi_i}$.  Each party performs according to his/her
local data a unitary transformation $ \ket{0}\bra{0} + e^{i
X_k}\ket{1}\bra{1}. $ Thus, the final  state is
\begin{equation}
\ket{\psi_f} = \frac{1}{\sqrt{2}}(\ket{0}+e^{i \sum_{k=1}^{N}
X_k}\ket{1}).
\end{equation}
The last party makes the same measurement as in task A. The
probability for the detection of state $2^{-1/2}(\ket{0} \pm
\ket{1})$, which we associate with the result $r=\pm1$, is given
by $P(\pm) = [1 \pm \cos(\sum_{k=1}^{N} X_k)]/2$. The expectation
value for the final answer $e_N=r$ is $ E(X_1,\ldots,X_N)= P(+) - P(-)$, and reads
$\cos(\sum_{k=1}^{N} X_k)$. The fidelity of  $e_N$, with
respect to $T_B$ is
\begin{eqnarray}
 F_{q,B} & = & \int_0^{2\pi}dX_1 \ldots \int_0^{2\pi}dX_N
 p_B(X_1,\ldots,X_N) \nonumber\\
  & & \times  T_B(X_1,\ldots,X_N)
 E(X_1,\ldots,X_N).
\end{eqnarray}
With the actual forms of $p_B$, $T_B$, and $E$, one gets $F_{q,B}
= \pi/4$, i.e., the protocol gives the correct value of $T_B$ with
probability $P_{q,B} = (1+\pi/4)/2 \approx 0.892$.

For both problems the classical fidelity $F_c$ or the probability
of success $P_c$ decreases exponentially with number $N$ of
parties to the value corresponding to a random guess of the result
of $T$, i.e. to the value achievable without any communication at
all. In contrast, $P_q$ remains constant for any $N$ and reaches 1
for task A, and $\approx 0.892$ for task B. The simple, one qubit
assisted quantum protocol clearly outperforms the best classical
protocols
without any shared multi-particle entanglement (!), utilizing only
the coherence properties of the transmitted qubit.

We implemented the quantum protocols for $N=5$ parties, using a
heralded single photon as the carrier of the qubit communicated
from one partner to the other \cite{coherence}. The qubit was
encoded in polarization. The computational basis, $``0"$ and
$``1"$, corresponds to horizontal $H$ and vertical $V$ linear
polarization, respectively. The data $X_k$ of each party was
encoded on the qubit via a phase shift using birefringent
materials. The last party performed a measurement in the
$2^{-1/2}(\ket{H}\pm\ket{V})$ basis in order to obtain the final
value $T$.

\begin{figure}[tb]
\includegraphics[width=86mm]{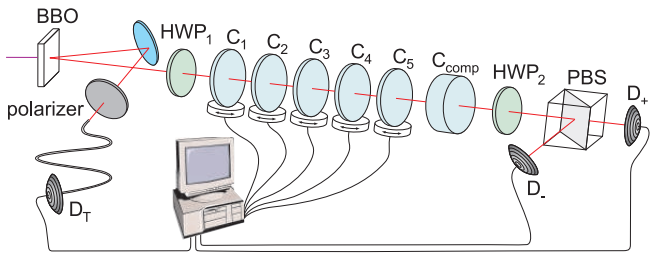}
\caption[Short caption]{\label{thelabel} Set-up for quantum CCP.
Pairs of orthogonally polarized photons are emitted from a BBO
crystal via the type-II SPDC process. The detection of one photon
as trigger at D$_\mathrm{T}$ indicates the existence of the other
one used in protocol. The polarization state is prepared with a
half-wave plate (HWP$_1$) and a polarizer, placed in the trigger
arm. Each of the five parties introduces a phase-shift by the
rotation of a birefringent YVO$_4$ crystal (C$_1$ to C$_5$). The
last party performs  the analysis of a photon-polarization state
using a half-wave plate (HWP$_2$) followed by a polarizing
beam-splitter (PBS).}
\end{figure}

The schematic set-up is shown in Fig. \ref{thelabel}. Photon pairs
are produced via spontaneous parametric down-conversion (SPDC).
The detection of one photon by detector D$_\mathrm{T}$ as a
trigger heralds the existence of the other one used in protocol.
The narrow gate window of 4 ns for observing the coincidence
detection between these two photons along with the single-count
rates of $\sim 140000$ s$^{-1}$ at the detectors D$_+$ and D$_-$
warrant that the recorded data are due to single photons only.
Type-II SPDC in 2 mm thick $\beta$-barium borate (BBO) crystal,
pumped by a single-mode laser diode (402.5 nm, 10 mW) is used,
emitting pairs of orthogonally polarized photons at $\lambda=805$
nm ($\Delta \lambda \approx 6$ nm). Filtering of the vertical
polarization of trigger photons by a polarizer, ensures that the
protocol photon has horizontal polarization initially. A half-wave
plate (HWP$_1$) transforms the state of the photon to
$2^{-1/2}(\ket{H} + \ket{V})$ as required in protocol.

For a fair comparison between the classical protocol and the
quantum protocol, no heralded events are discarded, even if the
detection of the protocol photon fails. In such a case it is still
allowed to guess the value of $T$. This works with probability
1/2, and  leads to very demanding experimental requirements for
unambiguous demonstration of the enhanced efficiency of
qubit-assisted CCP compared to its classical counterpart
\cite{galvao01}. In particular, high detection efficiency of the
heralded photons, i.e. high coincidence/single ratio for our
set-up, is essential.

In order to minimize the events where no photon was detected, the
yield of heralded photons was maximized by adopting an unbalanced
SPDC scheme. That means, we select a restricted spatial mode with
well defined polarization of the trigger photons by coupling them
into single-mode fibre behind a polarizer, whereas no spatial
filtering is performed on the protocol photons. With such
configuration we observed $\approx 5000$ trigger events per second
with $\approx 2400$ coincident events per second of protocol
detections, i.e. an overall detection efficiency of $\approx
0.48$, close to the limit given by the detector efficiency of the
avalanche photodiodes used, which was about $55\%$ for our
operating wavelength.

The individual phase shifts of parties are implemented by rotating
200 $\mu$m thick Yttrium-Vanadate (YVO$_4$) birefringent crystals
(C$_i$) along their optic axis, oriented perpendicularly to the
beam. An additional YVO$_4$ crystal (C$_{\rm comp}$) compensates
dispersion effects. To analyze the polarization state of photons
in the desired basis, a half wave-plate (HWP$_2$) followed by
polarizing beam-splitter (PBS) is used.

The protocols were run many times, to obtain sufficient
statistics. Each run took about one second. It consisted of
generating a set of pseudorandom numbers obeying the specific
distribution, subsequent setting of the corresponding phase shifts
by the rotations of YVO$_4$ crystals, and opening detectors for a
collection time window $\tau$. The limitation of communicating one
qubit per run requires that only these runs, in which exactly one
trigger photon is detected during $\tau$, are selected for the
evaluation of the probability of success $P_{exp}$. To maximize
the number of such runs, $n$, the length of $\tau$ was optimized
to 200 $\mu$s assuming Poissonian photon-number distribution of
SPDC photons.

In order to determine the probability of success from the data
acquired during the runs we have to distinguish the following two
cases. First, the heralded photon is detected, which happens with
probability $\eta$ given by the coincidence/single ratio. Then,
the answer on the value of the function $T$ can be based on the
measurement result. However, due to experimental imperfections in
the preparation of the initial state, the setting of the desired
phase shifts and the polarization analysis, the answer is correct
only with probability $\gamma$, which must be compared with the
theoretical limits given by $P_{q,A}$ and $P_{q,B}$ for the task A
and B, respectively. Second, with the probability $1-\eta$ the
detection of the heralded photon fails. Forced to make a random
guess, one gives the correct answer in half of the cases. This
leads to an overall success probability
$P_{exp}=\eta\gamma+(1-\eta)0.5$, or a fidelity of $F_{exp}=
\eta(2\gamma -1)$.

\begin{figure}[tb]
\includegraphics[width=86mm]{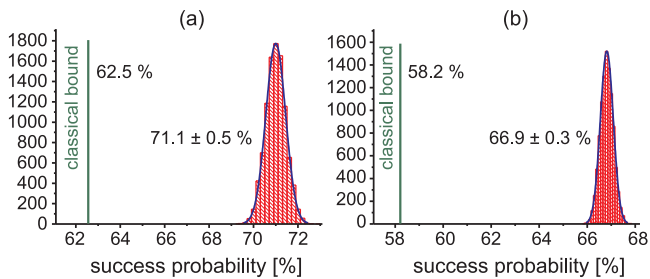}
\caption[Short caption]{\label{thelabel1} Histograms of measured
quantum success probabilities (a) for the task A and (b) for the
task B. The bounds for optimum classical protocols are displayed
as well.}
\end{figure}

Due to a finite measurement sample, our experimental results for
the success probability are distributed around the value $P_{exp}$
as shown in Fig. \ref{thelabel1} for both tasks. The width of the
distribution is interpreted as the error in the experimental
success probability. For task A we obtain a quantum
success probability of $P_{exp,A} = 0.711 \pm 0.005$.
The bound $P_{c,A} = 5/8$ for the optimum classical protocol is
violated by 17 standard deviations. For the task B we reached
$P_{exp,B} = 0.669 \pm 0.003$, whereas the classical bound is
$P_{c,B} \approx 0.582$. The violation is by 29 standard
deviations, \cite{fidelity}. Table \ref{tab} summarizes the
relevant experimental parameters $n$, $\eta$ and $\gamma$ for both
tasks.
\begin{table} [tb]
\caption{\label{tab}Experimental parameters}
\begin{ruledtabular}
\begin{tabular}{cccc}
 & $n$ & $\eta$ & $\gamma$ \\
task A & 6692 & $0.452\pm0.010$ & $0.966\pm0.003$\\
task B & 18169  & $0.471\pm0.006$ & $0.858\pm0.004$ \\
\end{tabular}
\end{ruledtabular}
\end{table}

In conclusion, we have proven and experimentally demonstrated the
superiority of quantum communication over its classical
counterpart for distributed computational tasks by solving two
exemplary CCP's. For nontrivial CCP's, where the
input from all the partners is required in order to obtain a
non-random final result, the best classical fidelity goes
exponentially to 0 with increasing number $N$ of partners. Yet, the
fidelity stays constant and independent on $N$ for our single
qubit assisted protocols.

In our experimental realization we have reached
higher-than-classical performance even when including all
experimental imperfections of state-of-the-art technologies. Thus,
by successfully performing fair and real comparison with the
classical scenario with present-day technology we clearly
illustrate the potential of the implemented scheme in real
applications of multi-party quantum communication. Most
importantly, our method gives a generic prescription to simplify
multi-party quantum communication protocols. For example,
multi-party secret-sharing protocols employing multi-qubit
GHZ-states and local operations only, can now be directly
transformed to single-qubit applications, thereby significantly
enhancing their applicability \cite{secret sharing}.

M.\.{Z}. was supported by Profesorial Subsidy of FNP, and by MNiI
grant No PBZ-MIN-008/ P03/ 2003. This work was supported by the
DFG, EU-FET (RamboQ, IST-2001-38864), Marie-Curie program and
DAAD/KBN exchange program.


\begin{thebibliography}{99}
\bibitem{broadcasting}
Broadcasting of bits, i.e., communicating them publicly, does not
improve the success rate of the classical protocol.
\bibitem{coherence}One should not be tempted to exchange the heralded
single photons carrying the qubits by bright polarized pulses of
light. In such a case, a suitable polarization measurement of the
pulses reveals all the encoded input data of any party, i.e. two
bits for task A and arbitrarily many for task B. Thus, the
communication restriction to $N-1$ bits is violated! The
attenuation of pulses to single-photon level does not help either.
The efficiency of the protocol would be significantly lowered, as
the attenuation causes a lot of non-detection events, forcing one
to guess the answer most of the time (see description of
experiment in main text).
\bibitem{fidelity}
Expressing the final results in terms of fidelities, we obtain
$F_{exp,A} = 0.421 \pm 0.010$ for task A, and $F_{exp,B} = 0.337 \pm 0.006$ for task B. The best
classical protocol reaches $F_{c,A}=0.25$ for A and
$F_{c,B}\approx0.164$ for B.
\end{thebibliography}
\end{document}